\documentclass[a4paper,twocolumn,aps,prb]{revtex4}

\newcommand{\eq}[1]{\begin{equation}#1\end{equation}}
\newcommand{\dd}{\mathrm{d}}
\newcommand{\ee}{\mathrm{e}}
\usepackage{amsmath}
\usepackage{graphics}
\usepackage{graphicx}
\usepackage{psfrag}

\begin{document}

\title{Fano resonances and entanglement entropy}


\author{Viktor Eisler}
\affiliation{Niels Bohr Institute, University of Copenhagen, Blegdamsvej 17,
DK-2100 Copenhagen \O, Denmark}
\author{Savannah Sterling Garmon}
\affiliation{Chemical Physics Theory Group, Department of Chemistry and Center for Quantum Information
and Quantum Control, University of Toronto, 80 St. George Street, Toronto, Ontario, Canada M5S 3H6}

\date{\today}

\begin{abstract}
We study the entanglement in the ground state of a chain of free spinless fermions with a single
side-coupled impurity. We find a logarithmic scaling for the entanglement entropy of a segment neighboring
the impurity. The prefactor of the logarithm varies continuously and contains an impurity contribution
described by a one-parameter function, while the contribution of the unmodified boundary enters additively. 
The coefficient is found explicitly by pointing out similarities with other models involving interface
defects. The proposed formula gives excellent agreement with our numerical data. If the segment has an open
boundary, one finds a rapidly oscillating subleading term in the entropy that persists in the limit of large
block sizes. The particle number fluctuation inside the subsystem is also reported.  It is analogous with the
expression for the entropy scaling, however with a simpler functional form for the coefficient.

\end{abstract}

\maketitle

\section{Introduction}

The entanglement properties of many-body systems have attracted considerable
attention and have become the topic of a large number of studies in the last decade
\cite{Amico_rev,CCD_intro}. In particular, the study of a suitable measure of entanglement
such as the entanglement entropy was triggered by the need to understand
ground state properties that lead to the emergence of area laws \cite{Eisert_rev}.
This keyword refers to a rather generic property of ground states of bipartitioned systems
in which the entropy $S$ of a subsystem scales as the number of contact points with the environment.
The most notable and well understood counterexamples are represented by one-dimensional critical
quantum systems with an underlying conformal symmetry where the area law is violated
in the form of universal terms, scaling as the logarithm of the subsystem size \cite{CC_rev}.
In the translationally invariant case this can be written as
\eq{S =\frac{c}{3}\ln L + k_0
\label{eq:enthom}}
with $c$ being the central charge of the conformal field theory (CFT) and $k_0$
a non-universal constant.
\par
The presence of impurities has interesting effects on the entanglement entropy
\cite{Affleck_rev}. An example is given by critical lattice models
where single defective links separate the two subsystems. This can lead to a modified
prefactor for the logarithm
varying continuously with the defect strength in models with free fermions
\cite{Peschel05,Levine/Miller08,Igloi/Szatmari/Lin09,EP10}, while for interacting electrons
the defect either renormalizes to a cut or to the homogeneous value in the $L \to \infty$
scaling limit \cite{Levine04,Zhao/Peschel/Wang06}. In the more general context of a conformal
interface separating two CFTs the effective central charge has been calculated recently
and was shown to depend on a single parameter \cite{SS08}.
\par
Another broad and intensively studied class of impurity problems is related to the Kondo effect\cite{Hewson}.
From the viewpoint of block entropies a spin-chain version of the Kondo model has been
studied \cite{SCLA07a,SCLA07b} using density matrix renormalization group (DMRG) methods
\cite{DMRG_book,Schollwoeck_rev}. Here a different geometry was considered with an impurity spin
coupled to one end of a finite chain. The induced change in the entropy varies
between zero and $\ln 2$ and its qualitative behavior for various system sizes and
coupling values can be described in terms of an impurity valence bond picture.
\par
A counterpart of the Kondo effect can be found in a single impurity model introduced by Anderson
\cite{Anderson61}.
A similar, exactly solvable model was studied independently by Fano \cite{Fano61} where the
interaction of a discrete state with a continuum of propagation modes leads to scattering resonances.
The exact form of these resonances is controlled by the couplings between the modes.
Here we consider a simple geometry where an impurity is side-coupled to a linear chain of
electron conduction sites. This setting can also be realized experimentally with gated semi-conductor
heterostructure quantum dots. In accordance with theoretical predictions \cite{Kang01}, Fano resonances
were detected as dips in the conductance measurements at low temperatures\cite{Kobayashi04}.
Despite the large amount of theoretical and experimental work on Fano resonances
\cite{Fanores_rev}, the question how they affect entanglement properties is still unanswered.
\par
In the present paper we address this question by investigating the simplest model capturing the main
features of the Fano resonance. The Fano-Anderson model \cite{Mahan} is described by a free fermion
Hamiltonian, thus standard techniques are available for the study of its entanglement properties \cite{PE_rev}.
We find that the entropy scaling of a block neighboring the impurity is of the form
(\ref{eq:enthom}) with a prefactor $c_{\mathrm{eff}}$ which decreases monotonously as the parameters
are tuned towards the Fano resonance and depends only on a well-defined scattering amplitude.
The numerical analysis leads to the same functional form of $c_{\mathrm{eff}}$ that
has recently been derived for simpler fermionic models with interface defects \cite{EP10}
and also seems to be closely related to the one found for conformal interfaces \cite{SS08}.
If the subsystem contains an open boundary we find a rapidly oscillating subleading term in
the entropy that persists even in the large $L$ limit.
\par
In the following we first introduce the model and the geometries considered,
along with the methods used to calculate the entropy. In Section \ref{sec:inf}
we treat an impurity attached to an infinite hopping model, determine
the correlation matrix and present results on the scaling of the entropy as well
as the particle number fluctuations. The effects of an open boundary will be detailed
in Sec. \ref{sec:sinf} followed by our concluding remarks in Sec. \ref{sec:concl}. 
The derivation of an asymptotic form of the correlation functions and some of
the necessary formulae for the semi-infinite case are presented in two Appendices.

\section{Model and methods \label{sec:mm}}

We consider a model of non-interacting spinless electrons described by a 1D tight-binding chain
that is coupled to an impurity.  The impurity is represented by an additional site from which tunneling events
can only occur to a specific site $n_0$ of the chain.
The Hamiltonian is
\eq{
\begin{split}
H= &-\frac{t}{2}\sum_{n}(c^\dag_n c_{n+1} + c^\dag_{n+1} c_{n})
+\epsilon_0 \sum_n c^\dag_n c_n \\
&-g \, (c_{n_0}^\dag d + d^\dag c_{n_0}) + \epsilon_d d^\dag d
\end{split}
\label{eq:ham}}
where $c_n$ and $d$ are the fermion annihilation operators along the chain and at the
impurity site, respectively, while $t$ and $g$ are the corresponding tunneling strengths.
The potential $\epsilon_0$ sets the filling of the chain and $\epsilon_d$ is the
site-energy of the impurity.
%
\begin{figure}
\center
\includegraphics[scale=0.6]{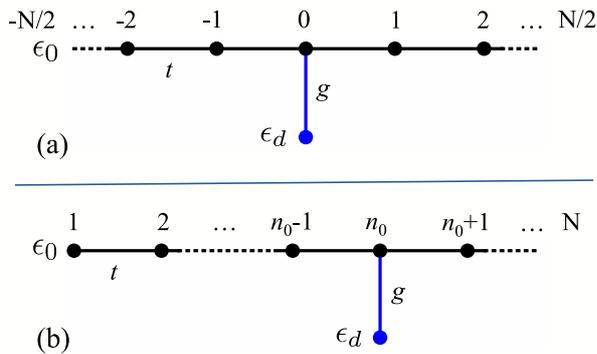}
\caption{(color online) Model geometry for (a) the infinite case and (b) the semi-infinite case.}
\label{fig:geom}
\end{figure}
\par
We will consider two different geometries:
\begin{itemize}
\item {in the \emph{infinite geometry} the $N+1$ sites are indexed as $n=-N/2, \dots, N/2$
while the impurity is located at the center of the chain ($n_0=0$) and we take periodic boundary conditions
}
\item {in the \emph{semi-infinite geometry} one has $N$ sites with $n=1, \dots, N$ while $n_0$ is arbitrary and we have open boundary conditions}
\end{itemize}
The two geometries are depicted in Fig. \ref{fig:geom}. In the following we set $t=1$.
After a Fourier transformation one has
\eq{H=\sum_q \epsilon_q c_q^\dag c_q + \sum_q t_q (c_q^\dag d + d^\dag c_q)
+ \epsilon_d d^\dag d 
\label{eq:hamq}}
where $\epsilon_q = \epsilon_0 -\cos q$. The couplings and the allowed wavenumbers are
$t_q = -\frac{g}{\sqrt{N+1}}$ with $q =\frac{2\pi}{N+1}n$ for the infinite and
$t_q = -g\sqrt{\frac{2}{N+1}}\sin n_0 q$  with $q =\frac{\pi}{N+1}n$ for the semi-infinite case,
respectively, where $n$ runs over the corresponding site-indices. The Fermi-level $\epsilon_{q_F}$
is set to zero by applying a potential $\epsilon_0 = \cos q_F$.
%
%
%
\par
The Hamiltonian (\ref{eq:hamq}) is known as the Fano-Anderson model and is diagonalized by
introducing new fermionic operators $f_q$ by the transformation \cite{Mahan}
\eq{d = \sum_q \nu_q f_q \quad , \quad c_q = \sum_{q'}\mu_{qq'}f_{q'}
\label{eq:munu}}
where
\eq{\nu_q = \frac{t_q}{\eta_{-}(\epsilon_q)} \quad \quad
\eta_{\pm}(\epsilon_q)= \epsilon_q-\epsilon_d-\Sigma(\epsilon_q \pm i\delta)
\label{eq:nu}}
\eq{\mu_{qq'} = \delta_{qq'} - \frac{t_q \nu_{q'}}
{\epsilon_q-\epsilon_{q'} + i\delta}
\label{eq:mu}}
with the self-energy term defined as
\eq{\Sigma(\epsilon_q \pm i\delta)=\sum_{q'}\frac{t_{q'}^2}
{\epsilon_q-\epsilon_{q'} \pm i\delta} .
\label{eq:sigma}}
The infinitesimal terms $\pm i\delta$ are needed to regularize the sum.
Note, that the number of different $q$ values equals the number of all sites,
including the impurity. In the limit $N \to \infty$ one has a continuum of the
unmodified single-particle spectrum $\epsilon_q$ of the chain while additional
bound states can emerge as real solutions $\epsilon$ of the equation
\eq{\epsilon-\epsilon_d-\Sigma(\epsilon)=0 \, .
\label{eq:bseq}}
\par
Our aim is to calculate the entanglement entropy $S=-\mathrm{Tr} (\rho \ln \rho)$
of a block of $L$ sites neighboring the impurity, where $\rho$ denotes the corresponding
reduced density matrix.
Since one is dealing with free fermions, this can be obtained through the eigenvalues
$\zeta_l$ of the correlation matrix $C_{mn}=\langle c_m^\dag c_n\rangle$ restricted to
the block $1 \le m,n \le L$ as \cite{Latorre/Riera_rev}
\eq{S = -\sum_l \zeta_l \ln \zeta_l - \sum_l (1-\zeta_l) \ln (1-\zeta_l) \, .
\label{eq:ent}}
\par

\section{Infinite geometry \label{sec:inf}}

We first consider the infinite geometry and calculate the matrix
$C_{mn}$ analytically; this result is in turn used to numerically obtain the entropy.
The particle number fluctuations are also considered where explicit
calculations are performed using an asymptotic form of the correlations.

\subsection{Correlation functions}

The correlations $\langle c_m^\dag c_n\rangle$ are calculated by first going over
to Fourier modes $c_q$ and subsequently applying the transformation (\ref{eq:munu}).
The expectation values to be evaluated are then trivial and read
$\langle f_q^\dag f_{q'}\rangle=\delta_{qq'} \chi(q)$ where in the ground state
$\chi(q)=1$ for the modes $\epsilon_q<0$ and zero otherwise. Moreover,
in the infinite geometry it was shown that Eq. (\ref{eq:bseq}) always gives
two real solutions with $\epsilon_{+}>1$ and $\epsilon_{-}<-1$ corresponding to bound states
above and below the band \cite{Tanaka06}. The energies $\epsilon_{\pm}$ are obtained
numerically by solving the quartic equation (\ref{eq:bseq}) with
\eq{\Sigma(\epsilon_{\pm})= \pm\frac{g^2}{\sqrt{\epsilon_{\pm}^2-1}}.
\label{eq:sigmapm}}
These modes $f_\pm$ have to be dealt with separately
in the transformation (\ref{eq:munu}) with the factors
\eq{\nu_{\pm}=\sqrt{N_{\pm}} \quad\quad \mu_{q \pm}=\sqrt{N_{\pm}}\frac{t_q}{\epsilon_{\pm}-\epsilon_d}}
where $N_{\pm}$ is set by the normalization condition $|\nu_{\pm}|^2 + \sum_q |\mu_{q \pm}|^2=1$.
Since $\langle f_-^\dag f_-\rangle=1$, one has a contribution $C^b_{mn}=C^b(m+n)$ to the correlations
coming from the lower bound state. Setting $t_q=-g/\sqrt{N+1}$ and taking $N\to\infty$ the sums are
replaced with integrals over the full band $\left[ -\pi,\pi \right]$ and can be rewritten as contour
integrals over the complex plane. Evaluating the residues one has
\eq{C^b(l) = \frac{g^2}{2\epsilon_-^2 - \epsilon_- \epsilon_{d} -1} \, \ee^{- l/\xi_{-}}
\label{eq:corrbinf}}
with $l=m+n$ for $m,n\ge0$ while the correlation length is defined by
$\xi_{-}^{-1}=\mathrm{arcosh}(-\epsilon_-)$.
%
%
\par
There are two contributions from the conduction band.
First we have the translationally invariant terms
\eq{C^0_{mn}= C^0(m-n)=\frac{\sin q_F(m - n)}{\pi (m - n)}
\label{eq:corr0inf}}
that give the correlations without the impurity arising from the $\delta_{qq'}$
term in (\ref{eq:mu}). The remaining term gives rise to the correlation contributions $C^1_{mn}=C^1(m+n)$;
these can be evaluated by applying the same methods used for determining $C^b(l)$. The calculation yields
the simple form
\eq{C^1(l) =  \int_0^{q_F} \frac{\dd q}{\pi} \sin \delta(q) \sin (l q + \delta(q))
\label{eq:corr1inf}}
where the scattering phases for $q>0$ are defined as
\eq{ \tan \delta(q) = \frac{\Sigma_q}{\epsilon_q - \epsilon_{d}}=
\frac{g^2}{ \sin q \left( \epsilon_d - \epsilon_{q} \right)}.
\label{eq:deltaqinf}}
Here $\Sigma_q=\mathrm{Im\,}\Sigma(\epsilon_q+i\delta)$
is the imaginary part of the retarded self-energy with the real part being zero.
\par
\begin{figure}[thb]
\center
\includegraphics[scale=.65]{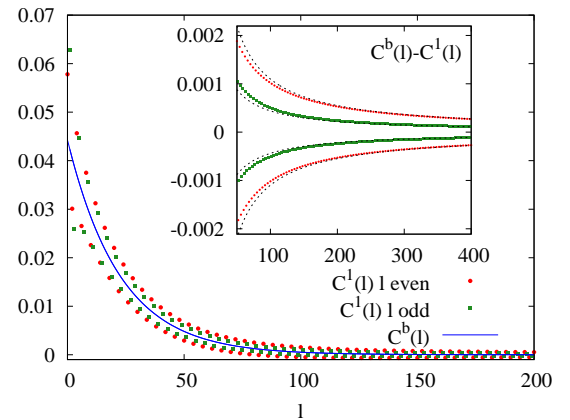}
\caption{(color online) Contributions to the correlations from the bound state $C^b(l)$ and
from the conduction band $C^1(l)$ for the parameter values $g=0.2$ and $\epsilon_d=-0.1$.
The inset shows the difference between these two contributions, compared to the asymptotic form (\ref{eq:corrfried})
represented by the dashed lines.}
\label{fig:corrfried}
\end{figure}

Collecting all the contributions, the correlation matrix is given by
\eq{C_{mn}=C^0_{mn}-C^1_{mn}+C^b_{mn}
\label{eq:corrinf}}
The integrals in (\ref{eq:corr1inf}) must be evaluated numerically and are shown on
Fig. \ref{fig:corrfried} together with the contribution of the bound state for some
fixed values of the parameters $g$ and $\epsilon_d$. The values of $C^1(l)$ are found
to oscillate around the exponentially decaying curve of $C^b(l)$. Since they appear
with opposite signs in (\ref{eq:corrinf}), the average value will cancel, corresponding
to the screening of the impurity induced localized state by the conduction electrons.
%
%
%
%
\par
The overall contribution  is shown in the inset of the Figure. A careful analysis
shows that the asymptotic behavior has the form of Friedel oscillations
\eq{C^b(l)-C^1(l) \approx C^F(l)=\frac{1}{\pi l} \sin \delta_F \cos(q_F l + \delta_F)
\label{eq:corrfried}}
where only the scattering phase $\delta_F = \delta(q_F)$ at the Fermi-level enters.
$C^F(l)$ is depicted by the dashed lines on the inset of Fig. \ref{fig:corrfried}
and shows a good agreement with the numerical data.
The derivation of Eq. (\ref{eq:corrfried}) is summarized in Appendix A. It relies on
an appropriate transformation of the integrals in (\ref{eq:corr1inf}) and is valid for
$l \gg l_0$ where the length scale $l_0$ is given in Eq. (\ref{eq:lnull}).
%
%

%

\subsection{Entanglement entropy}

The elements of the correlation matrix (\ref{eq:corrinf}) will be used to obtain
the entanglement entropy of a block numerically as described in Section \ref{sec:mm}.
In some simple cases one can already give an answer by looking at the limiting
form of the correlations. Taking $\epsilon_d \to \pm \infty$, the impurity site
will either be completely empty or completely occupied and therefore it becomes decoupled
from the rest of the chain. One has $\delta(q) \equiv 0$ while the contributions from the bound state
also vanish $C^b(l) = 0$, hence $C_{mn}=C^0(m-n)$ and the entropy will just be that of a homogeneous
hopping model given by (\ref{eq:enthom}) with $c=1$. Obviously, the limit of a vanishing coupling yields
the same behavior.
\par
On the other hand, one could take the limit $g \gg 1$ for very strong coupling.
This corresponds to $|\delta(q)| \to \pi/2$, that is, one has resonant
scattering at every wavelength. The ground state will correspond to a singlet
formed by the impurity and site zero that is otherwise decoupled from the rest of
the system. The bound state correlations contribute only on site zero
$C^b(l)=\delta_{l,0}/2$ while for sites $m,n>0$ one has $C_{mn}=C^0(m-n)-C^0(m+n)$,
which is the form for a semi-infinite chain. The entropy of this geometry
is known to scale logarithmically with a coefficient $1/6$, which is half of the value that appears in
the homogeneous case \cite{CC04}.
%
\begin{figure}[thb]
\center
\includegraphics[scale=.65]{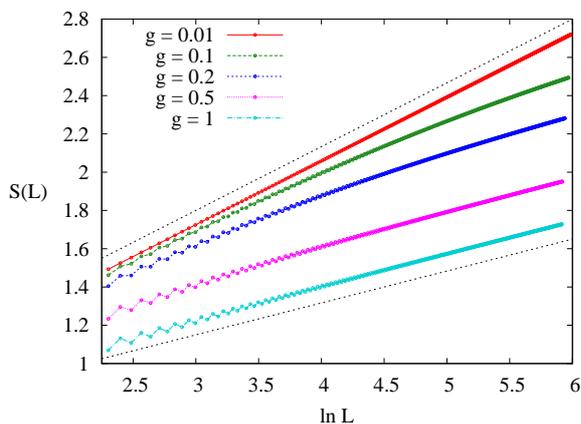}
\caption{(color online) Crossover in the entropy scaling in the resonant case $\epsilon_d=0$ at half
filling and for various coupling strengths $g$. The upper and lower-most dashed lines have slopes
$1/3$ and $1/6$, respectively, acting as guides to the eye.}
\label{fig:entcrossover}
\end{figure}
\par
The previous analysis shows that resonant scattering tends to decrease the
coefficient of the entropy scaling. However, in general a Fano resonance is
concentrated around a single wavenumber. This is expected to have the largest effect
on the asymptotic behavior if the location of the resonance coincides with
the Fermi-level $|\delta(q_F)|\to\pi/2$ yielding the condition $\epsilon_d \to \epsilon_{q_F}=0$.
We take this value of the site energy in Figure \ref{fig:entcrossover}, in which we plot the entropy
for a range of values for the coupling $g$.
We see that the entropy crosses over between the two limiting behaviors, giving a slope
of $1/3$ for small $L$ and $1/6$ for large $L$ when plotted against $\ln L$. Therefore,
taking the block size large enough the impurity behaves like a cut in the chain.
\par
This effect is reminiscent of the Kondo screening mechanism, where conduction
electrons form a cloud around a magnetic impurity. The Kondo effect has an associated length scale
\cite{Affleck09} that being the distance beyond which the impurity appears to be screened off.
In the present case a natural length scale also enters, which marks the asymptotic regime
of the correlations and could be expected to appear in the entropy scaling.
From Eq. (\ref{eq:lnull}) one has $l_0 \sim 1/g^2$ for $\epsilon_d=0$ and $q_F=\pi/2$, which
seems to be consistent with the location of the crossover moving toward larger $L$ in Fig.
\ref{fig:entcrossover} as the coupling $g$ decreases. In order to reliably verify the scaling of
the crossover length one would have to consider larger blocks. Unfortunately, the achievable sizes
were limited to $L \approx 400$ by the increasing numerical difficulty in evaluating the oscillatory
integrands for the matrix elements (\ref{eq:corr1inf}).
\par
For $L \gg l_0$ the screening cloud is effectively in a singlet state with the impurity electron and cuts the system
in two parts. The coefficient of the entropy therefore renormalizes to the value $1/6$ corresponding to a semi-infinite
chain with an open boundary. Note, that a similar renormalization behavior was found for \emph{interacting} electrons
in the Luttinger liquid regime bisected by a hopping defect \cite{Zhao/Peschel/Wang06,Levine04}.
\par
Our further numerical analysis shows that, for arbitrary parameter values and $L \gg 1$, the
entropy can be written in the form
\eq{S(L) = \frac{c_{\mathrm{eff}}}{3}\ln L + k
\label{eq:entropy}}
where the effective central charge $c_{\mathrm{eff}}$ varies continuously between the values
$1/2$ and $1$. The same behavior was found for free fermionic models with hopping defects
\cite{Peschel05,Igloi/Szatmari/Lin09} where the dependence of $c_{\mathrm{eff}}$ on the
defect strength was determined by data fits. Recently, Sakai and Satoh derived the same form as 
Eq. (\ref{eq:entropy})
for the case of two conformal field theories coupled through a conformal interface.  In their case they found
$c_{\mathrm{eff}}$ in a closed form with dependence only on a single scattering amplitude \cite{SS08}.
These conformal interfaces describe a discontinuity in the compactification radii of two bosonic CFTs,
which correspond to Luttinger liquids with unequal interaction parameters on the left and right
hand side \cite{CFTbook}.
\par
Although the result from \cite{SS08} is not directly applicable, a closely related form was recently found
to describe the case of free fermions with simple interface defects \cite{EP10}. We therefore attempted
to generalize these results to the present model. Motivated by Eq. (\ref{eq:corrfried}), we argue
that the asymptotic behavior of $S(L)$ and thus $c_{\mathrm{eff}}$ should depend only on the
scattering phase at the Fermi-level and a suitable scattering amplitude can be defined as $s=\cos \delta_F$.
Hence, we propose
\eq{c_{\mathrm{eff}}=\frac{1}{2} + \frac{6}{\pi^2}\int_{0}^{\infty}
u\left(\sqrt{1+(s/\sinh u)^2}-1\right)\dd u
\label{eq:ceffint}}
where $1/2$ comes from the unmodified boundary while the integral is, up to a
sign, the same as in \cite{SS08} and describes the contribution of the impurity.
\par
\begin{figure}[t]
\center
\includegraphics[scale=.65]{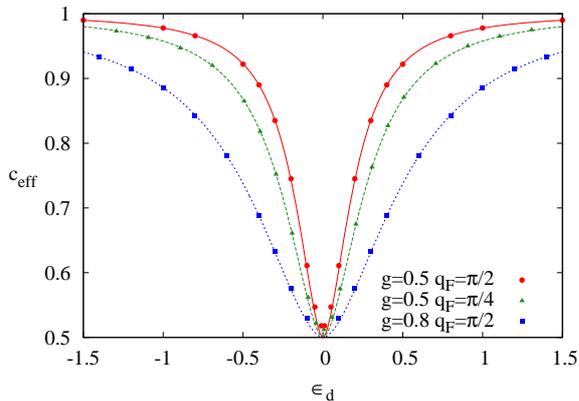}
\caption{(color online) Effective central charge, obtained from fits of the numerical data,
as a function of $\epsilon_d$ and for various values of $g$ and $q_F$.
The lines show the analytical form Eq. (\ref{eq:ceffint}) evaluated for the corresponding
values of the parameter $s$.}
\label{fig:ceffint}
\end{figure}

Figure \ref{fig:ceffint} shows the resulting $c_{\mathrm{eff}}$ obtained
by fitting (\ref{eq:entropy}) on different $S(L)$ data sets
together with (\ref{eq:ceffint}) evaluated as a function of $\epsilon_d$
for various fixed values of the coupling $g$ and filling $q_F$. One has excellent
agreement with the conjectured analytical form. The only visible deviations
arise for $\epsilon_d \approx 0$ which can be understood through the crossover
phenomenon shown on Fig. \ref{fig:entcrossover}. In this parameter regime the asymptotic
behavior sets in only for larger $L$ and one has considerable finite-size corrections.
%
%
\par
The integral (\ref{eq:ceffint}) depends on the model parameters only through the variable
\eq{s^2=\frac{\epsilon_d^2 \sin^2 q_F}{\epsilon_d^2 \sin^2 q_F + g^4}
\label{eq:intpar}}
Interestingly, the above formula \emph{exactly} coincides with the transmission coefficient
of the Fano-Anderson model \cite{Torio04} with the Fermi-energy set equal to zero.
Therefore, the effective central charge is directly linked to a simple physical quantity that,
in experimental realizations, is obtainable via conductance measurements at very low temperatures.
Although the description of Fano resonances measured in various nanostructure experiments
\cite{Fanores_rev} typically require more detailed impurity models, the simple relation between
$c_{\mathrm{eff}}$ and the transmission coefficient might still survive in some parameter regime.
This could open up the possibility of an indirect measurement of the leading term in the entropy.
\par
It should be pointed out that the impurity integral in $c_{\mathrm{eff}}$ has been
obtained \emph{analytically} for a transverse Ising chain in a DMRG geometry
\cite{EP10}. The calculation relies on a correspondence between transfer matrix
spectra of classical 2D Ising models and the single-particle eigenvalues of $\mathcal{H}$
in the reduced density matrix $\rho \sim \ee^{-\mathcal{H}}$ of the quantum chain. This analogy
is well-known in the homogeneous case \cite{PKL99,PE_rev} and can be extended by considering the
transfer matrix of a 2D strip with a defect line. In turn, an alternative representation
of the integral in (\ref{eq:ceffint}) via dilogarithm functions was obtained. In our present
model, however, a direct calculation of $c_{\mathrm{eff}}$ through the single-particle
eigenvalues seems to be a far more challenging problem.
%
\begin{figure}[thb]
\center
\includegraphics[scale=.65]{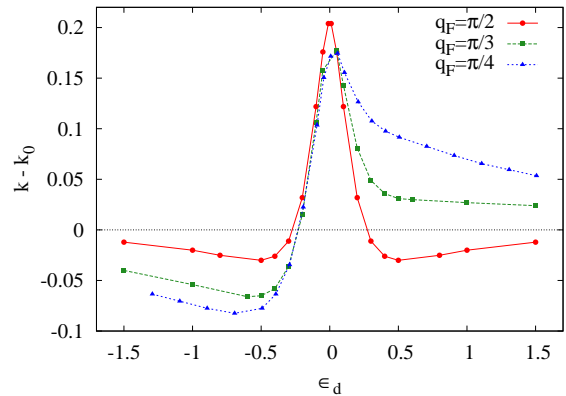}
\caption{(color online) Subleading term $k$ of the entropy as a function of $\epsilon_d$ for
different fillings and $g=0.5$. The homogeneous value $k_0$ has been subtracted for better comparison.}
\label{fig:kfits}
\end{figure}
\par
We also extracted the subleading contribution $k$ in (\ref{eq:entropy}) from our data
as shown in Fig. \ref{fig:kfits} for different values of the filling. Its value $k_0$ without
the impurity also depends on $q_F$ \cite{Jin/Korepin04} and was subtracted in the figure.
One has a peaked structure around $\epsilon_d=0$, however, unlike $c_{\mathrm{eff}}$ the constant
$k$ is in general not symmetric with respect to $\epsilon_d$. Instead, one has the relation
\eq{S(g,\epsilon_d,q_F)=S(g,-\epsilon_d,\pi-q_F)}
which is a direct consequence of the symmetry property Eq. (\ref{eq:corrsymm}) of the correlations
proven in Appendix A.

\subsection{Particle number fluctuation}

In the last part of this Section we will investigate the fluctuations in
the number of electrons $\hat{N}=\sum_{n=1}^L c^{\dag}_n c_n$ contained in the block
next to the impurity. Although being a simpler physical quantity, its scaling properties
were shown to be similar to that of the entanglement entropy in a variety of
1D quantum systems \cite{LeHur10} and, in the case of free fermions, also in arbitrary dimensions
\cite{Gioev/Klich06,Wolf06}. Therefore it is an interesting question whether this
connection persists for the present impurity problem.
\par
The particle number fluctuation is a simple quadratic function of the correlation
matrix and without the impurity is readily evaluated as \cite{Eisler/Legeza/Racz06}
\eq{\langle\Delta \hat{N}^2\rangle_0 = 
\mathrm{tr}\,{\bf C}^0({\bf 1}-{\bf C}^0) \approx \frac{1}{\pi^2} \ln L + k'_0}
with $k'_0=(\ln 2 +\gamma + 1)/\pi^2$. Note that here the leading-order logarithmic behavior
was seen to emerge from the large distance decay properties of the correlations.
Therefore, we will use the approximation in Eq. (\ref{eq:corrfried}) to write
${\bf C}\approx {\bf C}^0+{\bf C}^F$. Carrying out the traces, one finds that the additional
term $\mathrm{tr}\,({\bf C}^F)^2$ has a logarithmic contribution while
$\mathrm{tr}\,{\bf C}^F({\bf 1}-2{\bf C}^0)$ evaluates to a constant. The fluctuations then read
\eq{\langle\Delta \hat{N}^2\rangle =  \frac{\kappa_{\mathrm{eff}}}{\pi^2}\ln L + k'
\label{eq:fluctscale}}
with the prefactor given by $\kappa_{\mathrm{eff}}=(1 +\cos^2 \delta_F)/2$.
\par
\begin{figure}[thb]
\center
\includegraphics[scale=.65]{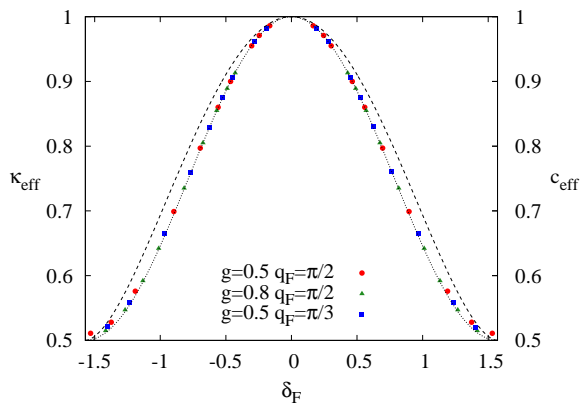}
\caption{(color online) Fitted coefficients $\kappa_{\mathrm{eff}}$ of the logarithmic term in the
particle number fluctuation for various values of $g$ and $q_F$, plotted against the scaling variable
$\delta_F$. The dotted line is the function $(1+\cos^2\delta_F$)/2. The dashed line shows $c_{\mathrm{eff}}$
as a function of $\delta_F$ for better comparison.}
\label{fig:fluctfits}
\end{figure}

The result is tested by comparing with the fitted values of $\kappa_{\mathrm{eff}}$ as shown
on Figure \ref{fig:fluctfits}. The agreement is very good with the only sizeable deviations appearing
in the region $|\delta_F| \approx \pi/2$ where the Fano-resonance is close to the Fermi-level and the
approximation (\ref{eq:corrfried}) breaks down for the numerically attainable block sizes.
Note, however, that in general the small distance deviations from the asymptotic form lead only
to a different numerical value of $k'$ as compared with the one obtained by evaluating
the subleading terms in the traces. Finally, one has to point out that, apart from the similar 
logarithmic scaling, the prefactor $\kappa_{\mathrm{eff}}$ of the fluctuations is given by a much simpler function
of the scattering amplitude than that governing the entropy.
%

\section{Semi-infinite geometry \label{sec:sinf}}

We now turn to the investigation of the semi-infinite model.
Our motivation behind studying a system with an open boundary is two-fold.
On one hand, we would like to cross-check our argument used for explaining
Eq. (\ref{eq:ceffint}) and show that the contributions in $c_\mathrm{eff}$ are additive.
In other words, we want to test our expectation $\tilde c_\mathrm{eff} = c_\mathrm{eff}-1/2$
for the semi-infinite effective central charge $\tilde c_\mathrm{eff}$ of a subsystem with an open end.
On the other hand, boundaries were shown to induce interesting subleading behavior
of the entropy for pure chains\cite{Laflo06}, therefore it is interesting to extend
these investigations to the present impurity problem.

\subsection{Correlation functions}

The main difference from the infinite case is that the couplings $t_q$ are now dependent on the
wavenumber. In principle the calculations are very similar, leading to slightly more
lengthy expressions that are therefore summarized in Appendix B. However, there is
an additional feature which leads to the emergence of bound states in the continuum
(BIC) which were studied before \cite{Longhi07,Tanaka07}. We will show how these BIC
naturally enter the problem at the correlation function level.
\par
Our main interest is to determine the entropy of a block with $L=n_0-1$ sites located
on the left hand side of the impurity. Apart from the usual bound state contribution,
the correlations from the conduction band now include
\eq{
\tilde C_{mn}= 2\int_{0}^{q_F} \frac{\dd q}{\pi} \, \mathcal{R}(q) \sin qm \sin qn
\label{eq:corr1sil}}
with $m,n < n_0$ and we have defined the function
\eq{\mathcal{R}(q)=\frac{(\epsilon_q - \epsilon_d)^2}
{(\epsilon_q - \epsilon_d -\Delta_q)^2+\Gamma_q^2}
\label{eq:resfun}}
where $\Delta_q$ and $\Gamma_q$ are the real and imaginary parts of the
self-energy (\ref{eq:sigmaqsi}), respectively, and are related to the parameter 
$\Sigma_q=\mathrm{Im\,}\Sigma(\epsilon_q+i\delta)$ of the infinite system as
\eq{
\Delta_q = \Sigma_q \sin 2n_0q \quad , \quad
\Gamma_q = \Sigma_q \,2\sin^2 n_0q \, .
\label{eq:reimq}}
Note that the term $C^0_{mn}$ does not now appear in the correlations.
\par
It is instructive to analyze the main features of the integral in Eq. (\ref{eq:corr1sil}). 
Taking $|\epsilon_d|\to\infty$ one has $\mathcal{R}(q) \equiv 1$ and the integral reproduces
the correlations of a pure semi-infinite chain.
For very strong coupling $g \to \infty$ one has $\Sigma_q \to \infty$ and therefore
both of the parameters in (\ref{eq:reimq}) take very large values apart from
a discrete set defined by the condition $\tilde q_n = n \pi / n_0$, $n=1,\dots n_0$ where
$\Delta_q = \Gamma_q = 0$. Therefore, the function $\mathcal{R}(q)$ will become sharply
peaked around these $\tilde q_n$ values and formally one can substitute
\eq{
\int \frac{\dd q}{\pi}\mathcal{R}(q) \to \frac{1}{n_0} \sum_{\tilde q_n}
\label{eq:intsum}}
where the factor $1/n_0$ is needed for proper normalization. Setting $n_0=L+1$,
the $\tilde q_n$ correspond exactly to the allowed wavenumbers in a \emph{finite}
chain of length $L$. Therefore the block is effectively decoupled by the impurity and
$\tilde C_{mn}$ reproduces the correlations in a finite segment.
%
\begin{figure}[thb]
\center
\includegraphics[scale=.65]{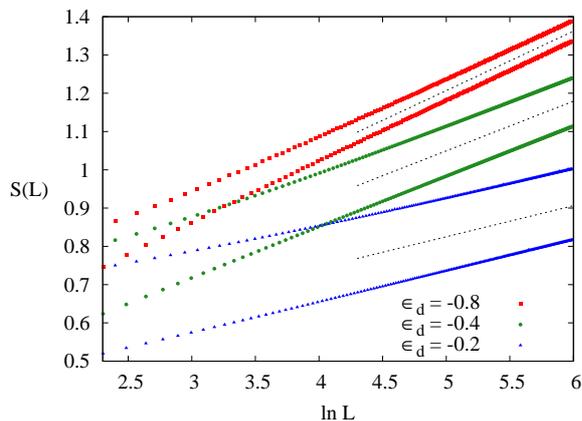}
\caption{(color online) Entropy scaling for a segment with an open boundary for different values
of $\epsilon_d$ at half filling and $g=0.5$. The dashed lines have corresponding slopes
$\tilde c_\mathrm{eff} = c_\mathrm{eff} -1/2$ and are shown for comparison.}
\label{fig:entsicomm}
\end{figure}
\par
The above argument breaks down for special $\tilde q_n$-values fulfilling
$\epsilon_{\tilde q_n}=\epsilon_d$ where $\mathcal{R}(q)$ becomes zero and the
corresponding peak is missing. However, this is exactly the condition for the
existence of the BIC. It has been shown that these states have non-vanishing amplitude
only at the impurity site and inside the segment where they reproduce, up to normalization,
the eigenstates of a chain of length $L=n_0-1$ \cite{Longhi07}. For $g \to \infty$ even the
normalization becomes exact and the contribution of the missing peak is therefore reincluded this way.
\par
While for these special values of $\epsilon_d$ the BIC are exactly located at $\tilde q_n$, the resonances of
$\mathcal R(q)$ gradually shift towards the wavenumbers of the decoupled block as the coupling
to the impurity becomes larger. Hence, their contribution to the entropy is expected to diminish.

\subsection{Entanglement entropy}

The numerical study of the entropy is now carried out using the exact form of the
correlations $\tilde C_{mn}+\tilde C^b_{mn}$. The scaling of the entropy at half filling is shown on Fig. \ref{fig:entsicomm}
for various $\epsilon_d$ values. In agreement with our expectations, the slope of the curves
coincides well with $\tilde c_\mathrm{eff} = c_\mathrm{eff}-1/2$ as illustrated by the
dashed lines. Additionally, one observes large amplitude oscillations in the data
which seem to persist for large values of $L$. Although such oscillations were already pointed
out in case of quantum chains with a boundary, they were found to decay according to a power law and
therefore vanish in the limit $L \to \infty$ \cite{Laflo06}.
\par
A qualitative argument for this alternation can be given by considering the main
parameters (\ref{eq:reimq}) entering the integrals. They contain oscillatory functions of $q$
that, for large $n_0$, are expected to average out upon integration, yielding
$\bar \Delta_q=0$ and $\bar \Gamma_q=\Sigma_q$. These are just the values for the infinite case,
explaining the average behavior of the entropy. However, as seen before the Fermi level plays
an important role in the asymptotics and at $q=q_F$ the parameters assume values which differ from the average.
For half-filling one has $\Delta_{\pi/2}=0$ while $\Gamma_{\pi/2}=0$ for $n_0$ even and 
$\Gamma_{\pi/2}=2\Sigma_{\pi/2}$ for $n_0$ odd, and thus oscillates around the average $\Sigma_{\pi/2}$.
This effect could result in subleading corrections to the entropy that survive the $n_0 \to \infty$ limit.
%
\begin{figure}[thb]
\center
\includegraphics[scale=.65]{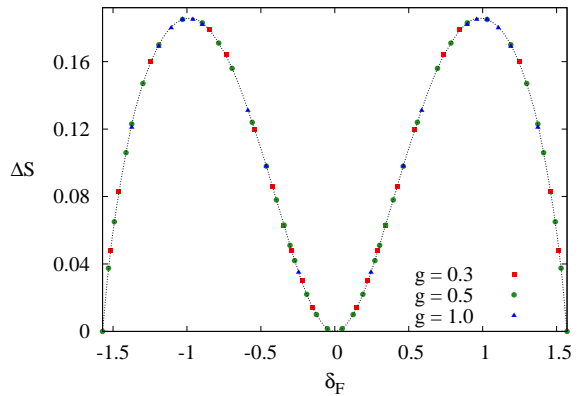}
\caption{(color online) Alternating term $\Delta S$ in the entanglement entropy at half-filling, as obtained
from data fits. The symbols correspond to different values of $\epsilon_d$ and $g$ and show a nice
collapse when plotted against the scaling variable $\delta_F$ of the infinite geometry.}
\label{fig:deltasfit}
\end{figure}
\par
To extract this term, we have fitted our data according to $S(L+1)-S(L)=\Delta S + a/L$.  This form, in general, gives a good
description of the alternating part. The results on $\Delta S$ are plotted on Fig. \ref{fig:deltasfit} against the
variable $\delta_F$. This choice is justified through our previous argument, where the only non-vanishing
parameter is given by $2\Sigma_{\pi/2}$ and thus the scattering phase of the infinite case is still expected to be a
relevant scaling variable. The data shows indeed a nice collapse to an interpolated scaling function which is shown
by the dotted line. It has a zero at $\delta_F=0$ corresponding to the pure semi-infinite chain where the alternating
term is known to vanish asymptotically \cite{Laflo06}. Note, that the symmetry under the exchange $\delta_F \to -\delta_F$
is valid also for $S(L)$ itself.
\par
\begin{figure}[thb]
\center
\includegraphics[scale=.65]{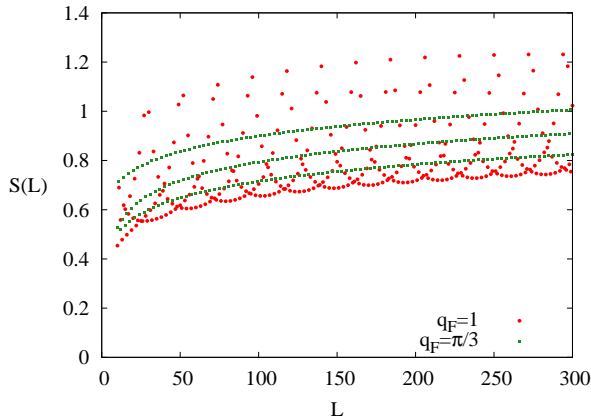}
\caption{(color online) Oscillating behavior of the entropy for a commensurate $q_F=\pi/3$ and
an incommensurate $q_F=1$ value of the filling. The parameters are $g= 0.5$ and $\epsilon_d= -0.8$.}
\label{fig:entsiincomm}
\end{figure}

The behavior of the entropy becomes more complicated when we choose another value of the filling.
This is illustrated on Fig. \ref{fig:entsiincomm} where we have compared $q_F=\pi/3$, corresponding to
one-third filling, with $q_F=1$ which gives an \emph{incommensurate} value for the filling. In the former
case the entropy curve splits into three parts, shifted from each other in a similar way as was
observed for half-filling on Fig. \ref{fig:entsicomm}. This corresponds to the three possible values
$\Delta_{\pi/3}$ and $\Gamma_{\pi/3}$ can take. However, in the incommensurate case the number of
distinct values is infinite, resulting in a highly irregular entropy behavior. Since $\pi/3 \approx 1.05$,
the effective central charges for the two cases are almost equal which explains why the averaged slope of
the curves is very similar. Nevertheless, for the small block sizes numerically achievable, the amplitude
of the oscillations has the same magnitude as the average entropy. Therefore, a quantitative understanding
of the subleading term would be clearly desirable and requires further investigation.
%
%
%

\section{Conclusions \label{sec:concl}}

We have studied the entanglement in the ground state of a single impurity problem described by the
Fano-Anderson model, focusing on the analysis of the effective central charge that appears in the
entropy scaling. We provided strong numerical evidence for the validity of a formula which
gives the functional form of $c_{\mathrm{eff}}$ for arbitrary parameter values of the model.
In particular, this formula depends only on a single parameter, given by Eq. (\ref{eq:intpar}), that has
a simple physical interpretation, namely it is the transmission coefficient of the impurity.
This, in turn, establishes a connection between the entanglement entropy and the
low temperature conductance of the quantum chain that can be measured in experiments
with quantum-dot nanostructures \cite{Fanores_rev}. Therefore, it would be interesting to generalize the
study of the block entropy for Anderson-type impurities, including on-site Coulomb interaction between
electrons of different spin, that give a more realistic model of the quantum-dots in experiments and yet
are still tractable with DMRG techniques.
\par
On the other hand, an exact analytical treatment of the present problem is still lacking. One could
follow the lines of Ref.\cite{EP10} and try to relate the reduced density matrix to the transfer matrix of
an appropriate 2D classical system. Another possible approach would be the direct calculation of the entropy
from the correlation matrix, which, in the pure case, requires knowledge of the eigenvalues of a Toeplitz
matrix \cite{Jin/Korepin04}. With an impurity one has a sum of Toeplitz and Hankel matrices,
which poses a much more difficult problem. However, the analysis of the correlations has led
to a simple asymptotic form, enabling us to calculate the particle number fluctuation that involves,
instead of diagonalization, only the traces of the correlation matrix.
\par
The investigation of the semi-infinite geometry has, on one hand, supported our general belief that the
leading contributions from both of the interfaces bordering the segment simply add up in the entropy.
On the other hand, we found a subleading term which shows an interesting oscillatory behavior. The
amplitude of the oscillations remains finite even asymptotically, an effect that has not been observed
for pure chains. Although some arguments were given to describe its main features, a proper understanding
could only be achieved through, analogous to the infinite case, a careful investigation of the asymptotic
correlations.
\par
Finally, it would be worth extending this study to the case where the impurity is located further apart
from the boundary of the subsystem. If this distance grows large, the effective central charge must
return to the value of a pure system and an interesting crossover behavior might emerge.

\begin{acknowledgments}
We thank Ingo Peschel and Michael Wolf for discussions. V.E. acknowledges financial support by 
the Danish Research Council, QUANTOP and the EU projects COQUIT and QUEVADIS. S.S.G. acknowledges
financial support from CQIQC.
\end{acknowledgments}

\appendix

\section{Asymptotic form of the correlations}

\begin{figure}[t]
\center
\includegraphics[scale=.4]{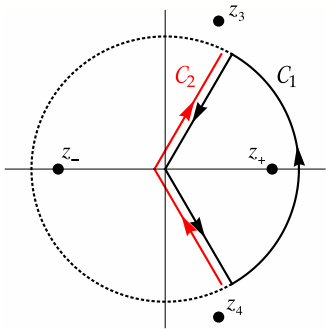}
\caption{(color online) Integration contours and poles of the integrand appearing in
(\ref{eq:corr1intc1}). The dashed line shows the boundary of the complex unit disk.}
\label{fig:intcont}
\end{figure}

The expression for the correlation functions involves the integrals in
Eq. (\ref{eq:corr1inf}) containing an oscillatory integrand. In the
limit $l \gg 1$ one could approximate it by considering the scattering phase
to be a constant. However, this method only works for slowly varying $\delta(q)$
functions which is not fulfilled by (\ref{eq:deltaqinf}) for $q \approx 0$
and $\epsilon_q \approx \epsilon_d$. It will be shown, that with a suitable
transformation of the integral one can extract the resonant contributions which
exactly reproduce the term $C^b(l)$ associated with the bound state.
To show this, we first rewrite (\ref{eq:corr1inf}) in the form
\eq{C^1(l)= \int_{-q_F}^{q_F}\frac{\dd q}{2\pi}
\frac{\Sigma_q^2 \cos ql + \Sigma_q (\epsilon_q-\epsilon_d)\sin ql}
{(\epsilon_q-\epsilon_d)^2 + \Sigma_q^2}
\label{eq:corr1infb}}
where we defined
\eq{
\Sigma_q = -\frac{g^2}{\sin q} = 
\begin{cases}
\mathrm{Im}\, \Sigma(\epsilon_q+i\delta), & \mbox{if } \quad q>0\\
\mathrm{Im}\, \Sigma(\epsilon_q-i\delta), & \mbox{if } \quad q<0
\end{cases}\, .
\label{eq:sigmaq}}
Note, that the denominator is just the product $\eta_+(\epsilon_q)\eta_-(\epsilon_q)$
which is strictly positive along the integration path in (\ref{eq:corr1infb})
but in general has four roots on the complex plane\cite{Tanaka06}.
Introducing the variable $z=\ee^{iq}$ the integration is now taken along an
arc of the complex unit circle which can be written as the sum of the two contours
shown in Fig. \ref{fig:intcont}.
%
%
After changing variables and using the symmetry under $z \to z^{-1}$ it takes the following form
\eq{
C^1(l)=\int\limits_{\mathcal{C}_1+\mathcal{C}_2} \frac{\dd z}{2\pi i} \,
\frac{4 g^2 z}{\Theta(z)}z^l
\label{eq:corr1intc1}}
with the fourth order polynomial
\eq{
\Theta(z)= z^4+2z(z^2-1)(\epsilon_d-\epsilon_0)+4g^2z^2-1
\label{eq:thetaz}}
\par
It is easy to show, that the polynomial has two real solutions $z_\pm = \pm \ee^{-1/\xi_\mp}$
which lie inside the unit disk and $z_+$ ($z_-$) is related to the lower (upper) bound
state solutions as $\epsilon_\mp = -(z_\pm+z^{-1}_\pm)/2=\mp \cosh \xi^{-1}_\mp$. The remaining solutions
form a complex-conjugate pair $z_{3}=z^*_{4}$ and lie outside the unit disk. The closed contour $\mathcal{C}_1$
therefore only encircles the pole at $z_+$ and the residue is evaluated as
\eq{
\underset{z=z_+}{\mathrm{Res \,\,}}\frac{4 g^2 z}{\Theta(z)}z^l = C^b(l)
\label{eq:res}}
where $C^b(l)$ is defined under Eq. (\ref{eq:corrbinf}). The remaining integral is taken along
the contour $\mathcal{C}_2$ which is parametrized as $z=\ee^{\pm i q_F} \ee^{-q'}$ and using
the analytic continuation of the scattering phase it reads
\eq{
-\mathrm{Re}\int_0^{\infty} \frac{\dd q'}{\pi} \sin \delta (q_F+iq')
\ee^{i (q_F+iq') l} \ee^{i \delta (q_F+iq')}
\label{eq:corr1intc2}}
Note, that $\delta (q_F+iq')$ varies smoothly and vanishes  for $q' \to \infty$. For $l \gg 1$
the integrand is rapidly decaying, thus one can approximate the phase around $q'=0$ as
$\delta (q_F+iq') \approx \delta_F + iq' \left. \frac{\dd \delta(q)}{\dd q}\right|_{q_F}$.
If the derivative is small, that is we are far away from the Fano resonance, one can set
$\delta (q_F+iq')=\delta_F$ and the integral (\ref{eq:corr1intc2}) can be trivially carried
out to yield the result in Eq. (\ref{eq:corrfried}). The derivative term is then used to define
a length scale
\eq{l_0 \sim  \left. \frac{\dd \delta(q)}{\dd q}\right|_{q_F}=
\frac{g^2(\sin^2 q_F - \epsilon_d \cos q_F)}{\epsilon_d^2 \sin^2 q_F + g^4}
\label{eq:lnull}}
which marks the regime $l \gg l_0$ where the approximation is valid. Note, that $l_0$ can only grow
large if the conditions $\epsilon_d^2 \ll g^2 \ll 1$ are satisfied.
\par
The contour integral can also be used to prove an exact symmetry property
of the correlations. We define the particle-hole transformed functions
$\bar{C}^b(l)$ and $\bar{C}^1(l)$ by performing the changes $\epsilon_d \to -\epsilon_d$
and $q_F \to \pi - q_F$. Now, one can show that the sum $C^1(l) + (-1)^l\bar{C}^1(l)$
can be written as the integral (\ref{eq:corr1intc1}) over the \emph{complete} unit circle
which is exactly evaluated as the sum of the residues at $z_+$ and $z_-$. Using the symmetry
property $\epsilon_+= -\bar{\epsilon}_-$, the latter pole contributes $(-1)^l \bar{C}^b(l)$
which, after reordering, yields the relation
\eq{
\bar{C}^b(l)-\bar{C}^1(l) = (-1)^l \left[ C^b(l)-C^1(l) \right].
\label{eq:corrsymm}}

\section{Semi-infinite formulae}

We present here some of the formulae which is necessary for the calculations
in the semi-infinite geometry. First, the self-energy function outside the band reads
\eq{
\tilde\Sigma(\epsilon_\pm)=\pm\frac{g^2}{\sqrt{\epsilon_\pm^2 - 1}}
\left[ 1-\left(| \epsilon_\pm |- \sqrt{\epsilon_\pm^2 - 1} \right)^{2n_0}\right]
\label{eq:sigmapmsi}}
and the bound state energies $\epsilon_+ > 1$ and $\epsilon_- < -1$ are obtained
numerically as the roots of
\eq{\epsilon_\pm-\epsilon_d-\tilde\Sigma(\epsilon_\pm)=0 \, .
\label{eq:bseqsi}}
Note, that $\epsilon_+$ ($\epsilon_-$) exists only for parameter values fulfilling
$\epsilon_d > 1-2 g^2 n_0$ ($\epsilon_d < -1+2 g^2 n_0$). On the other hand, using the definition
of the inverse correlation length $\epsilon_\pm = \pm \cosh \xi^{-1}_\pm$ the second term
in the parentheses in (\ref{eq:sigmapmsi}) can be written as $\ee^{-2n_0/\xi_\pm}$.
Therefore, if the impurity lies deep in the bulk of the chain with $n_0 \to \infty$ one has
$\tilde\Sigma(\epsilon_\pm) = \Sigma(\epsilon_\pm)$ and consequently the same solutions for
$\epsilon_\pm$ as in the infinite case.
\par
The contribution $\tilde C^b_{mn}$ from the lower bound state can be evaluated as
\eq{
\frac{g^2 \tilde N_-}{\epsilon^2_- - 1}
\begin{cases}
4\sinh^2 \frac{n_0}{\xi_-}\ee^{-(m+n)/\xi_-} & \text{if $m,n \ge n_0$} \\
4\sinh \frac{m}{\xi_-} \sinh \frac{n}{\xi_-}\ee^{-2n_0/\xi_-} & \text{if $m,n < n_0$}
\end{cases}
\label{eq:corrbsi}}
with the normalization factor
\eq{
\tilde N_- = \frac{\epsilon_-^2-1}
{2\epsilon_-^2 - \epsilon_- \epsilon_d -1 -2n_0 g^2 \ee^{-2n_0/\xi_-}}\, .
\label{eq:normsi}}
The limit $n_0 \to \infty$ gives $\tilde N_- = N_-$ and the right hand side of
(\ref{eq:corrbsi}) becomes $\ee^{-(m'+n')/\xi_-}$ with the shifted indices
$m'=|m-n_0|$ and $n'=|n-n_0|$, thus recovering again the result (\ref{eq:corrbinf})
for the infinite case.
\par
Finally, one needs the expression for the retarded self-energy inside the band
\eq{
\tilde\Sigma(\epsilon_q+i\delta)=-2g^2 \frac{\sin n_0 q}{\sin q} \ee^{i n_0 q}
\label{eq:sigmaqsi}}
which is an oscillatory function of the position $n_0$. This eventually leads to
the emergence of bound states in the continuum, as described in the text.
The scattering phases are defined as
\eq{
\tan \tilde\delta(q) = \frac{\Gamma_q}{\epsilon_q - \epsilon_d - \Delta_q}
\label{eq:deltaqsi}}
with $\Delta_q = \mathrm{Re\,}\tilde\Sigma(\epsilon_q+i\delta)$ and
$\Gamma_q = \mathrm{Im\,}\tilde\Sigma(\epsilon_q+i\delta)$.
\par
The band contributions
also have to be treated separately on either sides of the impurity.
On the right hand side, analogously to the infinite case, it can be expressed
in the form $\tilde C^0(m-n)-\tilde C^1(m+n)$ with
\eq{\tilde C^1(l)=\int_{0}^{q_F}\frac{\dd q}{\pi} \cos(ql+2\tilde\delta(q))
\label{eq:corr1sir}}
while the translationally invariant term  $\tilde C^0(m-n) = C^0(m-n)$
is unmodified. The correlations on the left hand side cannot be given in
that simple form and are analyzed in the text.



\end{document}